\documentclass{article}

\usepackage[final]{amlc_2023}

\usepackage{etoolbox}
\AtBeginEnvironment{tcolorbox}{\tiny}

\usepackage[utf8]{inputenc} % allow utf-8 input
\usepackage[T1]{fontenc}    % use 8-bit T1 fonts
\usepackage{hyperref}       % hyperlinks
\usepackage{url}            % simple URL typesetting
\usepackage{booktabs}       % professional-quality tables
\usepackage{amsfonts}       % blackboard math symbols
\usepackage{nicefrac}       % compact symbols for 1/2, etc.
\usepackage{microtype}      % microtypography
\usepackage{xcolor}         % colors
\usepackage{graphicx}
\usepackage{textcomp}
\usepackage[linesnumbered]{algorithm2e}
\usepackage{array}
\usepackage{tabulary}
\newcolumntype{K}[1]{>{\centering\arraybackslash}p{#1}}
\usepackage{diagbox}
\usepackage{caption}
\usepackage{relsize}
\usepackage{verbatim}
\usepackage{pbox}
\usepackage{tabularray}
\usepackage{xspace} % from christos, for the naming macros
\usepackage{paralist} % from christos, for compactitem

\usepackage{amsthm}
\usepackage{subcaption}
\usepackage{enumitem}
\usepackage{multirow}
\usepackage{tabularx}
\usepackage{new_defs}
\usepackage{algorithm,algorithmic}
\usepackage{tcolorbox}

\usepackage[para]{footmisc}
\usepackage{soul}

\usepackage{cleveref}

\usepackage{tikz}
\usetikzlibrary{trees}
\usetikzlibrary{positioning}
\usetikzlibrary{matrix}
\definecolor{myyellow}{RGB}{250, 246, 145}
\definecolor{myorange}{RGB}{245, 203, 130}
\definecolor{mygreen}{RGB}{211, 247, 188}
\definecolor{myblue}{RGB}{86, 180, 233}
\def\BibTeX{{\rm B\kern-.05em{\sc i\kern-.025em b}\kern-.08em
    T\kern-.1667em\lower.7ex\hbox{E}\kern-.125emX}}

\newtheorem*{definition}{Definition}

\usepackage{relsize}

\usepackage[colorinlistoftodos]{todonotes}

\newcommand{\para}[1]{\noindent\textbf{#1.}}

\theoremstyle{definition}

\definecolor{commentgreen}{rgb}{0,0.5,0}

\usepackage{colortbl}

\SetCommentSty{mycommfont}

\newcommand\method{\textsc{LakeGen}\xspace}

\title{\method: A LLM-based Tabular Corpus Generator for Evaluating Dataset Discovery in Data Lakes}

\author{
  Zhenwei Dai \\
   Amazon \\
   \texttt{zwdai@amazon.com} \\
   \And
   Chuan Lei \\
   Amazon \\
   \texttt{chuanlei@amazon.com} \\
   \AND
   Asterios Katsifodimos \\
   Delft University of Technology \\
   \texttt{akatsifo@amazon.com} \\
   \And
   Xiao Qin \\
   Amazon \\
   \texttt{drxqin@amazon.com} \\
   \And
   Christos Faloutsos \\
   CMU \\
   \texttt{faloutso@amazon.com}
   \And
   Huzefa Rangwala \\
   Amazon \\
   \texttt{rhuzefa@amazon.com} \\
}

\begin{document}

\maketitle

\begin{abstract}

How to generate a large, realistic set of tables along with joinability relationships, to stress-test dataset discovery methods? Dataset discovery methods aim to automatically identify related data assets in a data lake. The development and evaluation of such solutions for customers from a wide range of business domains, relies on diverse, high quality and domain-specific tabular benchmarks.
Large language models (LLMs) are trained on a wide variety of text data, which can provide a strong foundation of general and domain-specific knowledge. In this paper, we ask the question -- \textit{can we leverage LLMs to generate a tabular benchmark adequate for evaluating the dataset discovery solutions?} In particular, we focus on the task of finding joinable tables which is the cornerstone of virtually every dataset discovery method. Current corpora for evaluating dataset discovery methods  are mainly based on subsets of open data, and they suffer from three important issues: $i)$ they focus on very common and generic data types  (e.g., address, id, name, etc.);  $ii)$ they do not contain human-annotated column pairs; instead, practitioners synthesize ground truth using table splits (e.g., horizontal for table union search and vertical ones for joinability) and $iii)$ they do not focus on semantic column relationships.

% \huzefa{The above paragraph should preface on why dataset joinability is important? We miss that for a non-DB audience like AMLC}
%Done, please check!

In this paper we present \method, a method for creating and curating domain-specific datasets for training and evaluating dataset discovery methods, by using LLMs. 
\method makes the following contributions:
(a) \novel: it is the first to use ontologies, LLMs, as well as semantic perturbations for corpora generation.
(b) \general: we have used \method to curate three corpora for diverse domains, including finance, pharma, and healthcare with minimal effort.
(c) \effective: our experiments show that the generated domain-specific corpora pose multiple challenges to existing union search and joinability discovery methods, proving our intuition -- current datasets fail to evaluate these methods properly.
\end{abstract}

\vspace{-5mm}

% \textbf{All papers must include a minimum of 1-paragraph that discusses the Amazon customer problem of the topic presented and must state whether this is a short-term, long-term or existing impact.} Please pay special attention to the instructions after section 6, regarding the Customer Problem Statement. This statement must be included in your 8-page limit.

% {\bf Note that the reference section does not count towards the eight pages of content that are allowed.}

\section{Introduction}
\label{sec:intro}

% \christos{how about a rhetorical question: How can we generate a large, realistic set of tables, to stress test data-lake algorithms?} % chuan - the opening statement is added to the abstract.

Dataset discovery tasks are a critical component of modern data-driven workflows, where finding and integrating relevant datasets in large data lakes is essential for many data analytics and machine learning tasks~\cite{doan2012principles,ritchie2015methods}. Modern dataset discovery tasks such as the one of finding semantic joinable columns of tables are challenging as tables in a data lake may have different way of conveying the same information~\cite{dong2021efficient}. Being able to understand the tabular data by appropriately profiling the syntactic and semantic rich information becomes essential for the success of a data discovery system~\cite{dong2021efficient,fan2022semantics}. 

To evaluate the effectiveness of dataset discovery methods, researchers rely on corpora of datasets alongside ground truth (i.e., annotated column/table pairs) that are specifically curated for the purpose of dataset discovery. However, the existing tabular datasets \cite{lee2007etuner,koutras2021valentine,nargesian2018table,hulsebos2021gittables,tpcdi,gaulton2012chembl,magellandata} do not present the necessary semantic complexity, domain-specificity, and annotations for different discovery tasks, hindering their ability to evaluate the effectiveness and robustness of modern data discovery methods.

% \christos{How about we move the next paragraphs, into the 'related work' section - and here we just say: As we discuss in *related-work-section*, none of the existing methods satisfy all the requirements that the proposed \method satisfies - see Table~\ref{tab:related-works}} % chuan - I would avoid forward references. Also we need this paragraph to explicitly motivate why we need \method to generate semantically rich domain-specific datasets.

\para{Pitfalls of current corpora used in dataset discovery}
For instance, the recently published tabular corpus GitTables~\cite{hulsebos2021gittables} comprises open data, and its semantic data types cover the commonly used ones in open knowledge bases, such as addresses, age, names, etc. This lack of semantic complexity limits its ability to evaluate methods dealing with complex and diverse domain-specific data types.

Dataset discovery in scientific domains highlights the importance of handling complex data types, such as chemical structures, genome sequences, and astronomical data. Similarly, dataset discovery in healthcare often involves handling complex medical concepts and terminologies, such as ICD codes in the Systematized Nomenclature of Medicine Clinical Terms (SNOMED CT) \cite{donnelly2006snomed}. Such a domain-specificity is not included in GitTables. In addition, GitTables does not provide any ground truth annotations for dataset discovery. For example, it does not contain annotations of column pairs that can be used for joining tables. Smaller corpora such as TPC-H and TPC-DS~\cite{nambiar2006making} do offer ground-truth annotations, but they cover only \textit{primary key}/\textit{foreign key} (PK/FK) relationship, which can be easily identified using set similarity metrics over the data instances. This does not resemble complex relationships found in real-world data lakes. 

Since existing corpora used for dataset discovery offer a limited amount of human-annotated ground truth, recently proposed methods attempt to create synthetic relationships via table splitting  \cite{nargesian2018table,koutras2021valentine,fan2022semantics}. Vertical table splitting generates two tables that can be joined, and horizontal table splitting generates two unionable tables. However, those table splits fail to correctly evaluate dataset discovery methods, as they do not offer challenging joinability scenarios. Despite the use of na\"ive noise~\cite{lee2007etuner, koutras2021valentine} (e.g., typos, jitter, etc.) to make those matches more challenging to dataset discovery methods, these splits still lead to correspondences across columns (or tables), which do not reflect the real-world challenges pertaining to data discovery. 
% Due to privacy constraints, even large enterprises~\cite{zhang2023schema} do not have easy access to customers' domain-specific datasets for training and validating their own data discovery solutions due to issues of privacy or IP protection.

\para{LLMs as a source of semantic, and domain-specific datasets} At the same time, Large Language Models (LLMs), trained on massive amounts of data, perform well at a wide variety of tasks including text summarization, translation, text generation and many others. Recently, we have seen emerging efforts towards synthetic data generation by leveraging LLMs \cite{DBLP:journals/ijon/HernandezEACR22,borisov2023language}. Essentially, the off-the-shelf LLMs can be used as knowledge bases as they often capture high quality semantic relationships present in the training data. At the same time, it has been shown that factual knowledge can be recovered surprisingly well from the LLMs \cite{DBLP:conf/emnlp/PetroniRRLBWM19}. As we show in \Cref{sec:base_tables_generation}, LLMs score equally well in generating domain-specific datasets.

\definecolor{ChromeWhite}{rgb}{0.811,0.925,0.776}
\begin{table}[ht]
\centering
\label{tab:related-works}
\resizebox{\textwidth}{!}{%
\begin{tblr}{
  row{even} = {c},
  row{3} = {c},
  row{5} = {c},
  row{7} = {c},
  row{9} = {ChromeWhite,c},
  cell{1}{2} = {c=2}{c},
  cell{1}{4} = {c=3}{c},
  cell{1}{7} = {c=2}{c},
  cell{1}{9} = {c=2}{c},
  vline{2-11} = {1}{},
  vline{-} = {2-9}{},
  hline{1} = {2-10}{},
  hline{2-10} = {-}{},
}
& \textbf{{Semantic Data Types}}    &   & \textbf{{Ground Truth (joinable columns)}} &    &   & \textbf{{Targeted Problem}}   &   & \textbf{{Value contents}} & \\
\textbf{{Corpus}} & \textbf{Common}               & \textbf{Domain Specific} & \textbf{Split-based}    & \textbf{Human-annotated} & \textbf{PK/FK} & \textbf{Unionability}    & \textbf{Joinability} & \textbf{Synthetic}       & \textbf{Real-world} \\
Open Data \cite{nargesian2018table}            & $\checkmark$                  &                          & \cellcolor{green} $\checkmark$            & $\checkmark$             &                & $\checkmark$               &                      &                          & $\checkmark$        \\
ChemBL \cite{gaulton2012chembl}                & $\checkmark$                  &   $\checkmark$                       &                         &                          & $\checkmark$   &                            & $\checkmark$         &                          & $\checkmark$        \\
TPC-DI \cite{tpcdi}              & $\checkmark$                  &                          &                         &                          & $\checkmark$   &                            & $\checkmark$         & $\checkmark$             &                     \\
Magellan \cite{magellandata}             & $\checkmark$                  &                          &                         & $\checkmark$             &                &                            & $\checkmark$         &                          &                     \\
Valentine (WikiData) \cite{koutras2021valentine} & $\checkmark$                  &                          & $\checkmark$            & $\checkmark$             &                & $\checkmark$               & $\checkmark$         &                          &                     \\
GitTables \cite{hulsebos2021gittables}             & $\checkmark$                  &                          &                         &                          &                &                            &                      &                          & $\checkmark$        \\ \method{}
                      & $\checkmark$                  & $\checkmark$             & $\checkmark$            & $\checkmark$             & $\checkmark$   & $\checkmark$               & $\checkmark$         &                          & $\checkmark$        
\end{tblr}
}
\caption{\myTag{\method wins:} Existing corpora that are used for dataset discovery evaluation.}
\end{table}

\noindent This paper addresses the following question: \textit{Given a set of semantic data types that describe real-world entities, how can we generate realistic, domain-specific datasets and ground truth with column relationships, for dataset discovery scenarios?}

\begin{figure*}[t]
    \centering
    \includegraphics[width=.99\textwidth]{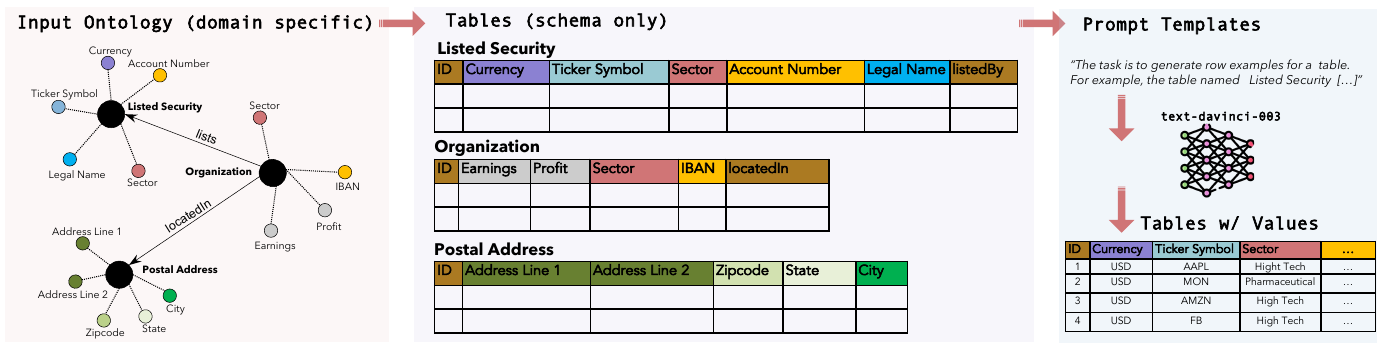}
    \caption{\myTag{\method overview:} Base-table generation using an ontology and a LLM.\vspace{-4mm}}
    \label{fig:ontology-to-table}
    % \vspace{-3mm}
\end{figure*}

\para{Approach overview} 
The approach of \method{} is to leverage the generative capabilities of LLMs to generate synthetic datasets that mimic the semantic complexity and diversity of real-world tabular data lakes. This paper presents \method{}, a dataset and ground truth synthesizer for data discovery tasks.
As seen in \Cref{fig:ontology-to-table}, a user (e.g., a client or a dataset discovery method developer) can make use of an ontology to choose the entities alongside their properties in order to generate tables that resemble a given data domain. In this example, we show a simplified subset of FIBO, a widely adopted financial-domain ontology ~\cite{fibo}. The ontology contains three entities and their data properties. In addition, it contains relationships (i.e., object properties) among those entities. \method{} starts by first converting the ontology into a set of table schemata with relationships across them. It then uses those schemata, to prompt a LLM to generate values and populate those schemata. By using the data types information from the ontology (or human annotations) as shown in \Cref{fig:table-to-benchmark}, after creating tables and their values, \method{} can  infer relationships across tables (e.g., joinability). For instance, if two tables contain a column of the "Account Number" semantic data type, \method{} can infer that those columns are joinable. After a set of perturbations (\Cref{sec:table_perturb}), \method{} generates a set of tables, populated with values sourced from LLMs, and ground truth inferred from the given ontology. The benchmark contains a set of tables alongside their relationships, and those can be used to evaluate a given data discovery (e.g., joinable, unionable, etc.) method.

% \para{Approach Overview} In short, as seen in \Cref{fig:ontology-to-table} \method{} works as follows: a user provides an existing ontology that describes object types and their relationships, as well as the semantic data types of the object properties. \method{} then uses the provided object types and their properties to generate a complex database schema: objects become tables, and their properties become table attributes. After that, the table schema needs to be filled with data. To this end, \method{} leverages a LLM to generate realistic rows for those tables. As seen in \Cref{fig:table-to-benchmark}, the matches between tables are established via the ontology: the semantic data types extracted from object properties (as well as their subtypes), create pairs of attributes that form a match. To make the matching more challenging, \method{} propose a number of table perturbation methods, to introduce dataset noise.

% \asterios{From CF: decide the contributions and give some numbers. Accurate or realistic makes it strong. Numbers of joins, Automatic, realistic and scalable. The realistic part is the the hard part. Find measures of realism and deliver on those. Green for checkmarks.}

\noindent This paper proposes \method, which has the following desirable properties:
% we make the following contributions.

% \huzefa{Novel is not a contribution. All contributions are novel. Should we say \method is 1. can take in ontologies as input and use LLM to populate base tables with minimal user input. 2. Can do semantic pertubations and 3. Works across multiple domains and generates.}
% Good points, I have "toned it down", keeping the contributions. 

\begin{compactitem}
\item \novel: \method is the first work to use ontologies to drive table schema generation and to utilize a LLM to populate these base tables, with minimal user input. Moreover, \method proposes semantic perturbations to introduce noise and increase complexity at the level of column names, table shape and column values.
\item  \general: \method can generate  domain-specific datasets
for diverse domains. In \Cref{sec:experiments} we generate three datasets for the domains of finance, pharma, and healthcare. These datasets present a variety of challenges for the dataset-discovery task.
\item \effective: \method illustrates the weaknesses of recent dataset discovery methods, as shown in our experimental study (\Cref{sec:experiments}).
\end{compactitem}

% \asterios{repeat the contributions to abstract \& conclusions. Crown figure should corroborate the contributions. We could have a histogram with financial, healthcase \& HR. Continuum creates issues to other}

\vspace{-4mm}
\section{The Column Joinability Problem}
\label{sec:problem}
\vspace{-3mm}

In this paper we are interested in the problem of generating domain-specific benchmarks for one critical dataset discovery task, namely joinability discovery. Joinability discovery is a form of schema matching~\cite{rahm2001survey} and is the process of identifying whether two columns from two different tables can be joined. We will represent two tables as $R$ and $S$, and their corresponding column sets $C_R$ and $C_S$. We formally define \textit{column joinability} of two columns $c_R \in C_R$ and $c_S \in C_S$ (denoted as $c_R \bowtie c_S$) as follows.

\begin{definition}[\textbf{Column Joinability}]
Two columns $c_R \in C_R$ and $c_S \in C_S$ are joinable if $c_R$ and $c_S$ can be joined through a mapping function $h$, which maps the values in $c_R$ to the values in $c_S$, i.e., $(c_R \bowtie_{h(c_R) = c_S \lor c_R=h(c_S)} c_S) \neq \emptyset$
\end{definition}

Note that if the mapping function $h$ is the identity function, joinability refers to exact-joinability (also known as \textit{equi-joins}) \cite{koutras2021valentine}. On the other hand, if we add two extra constraints to the definition of column joinability, namely $i)$ $c_S \subseteq c_R$ and $ii)$ $c_R$ is the primary key of relation $R$, then the joinability refers to primary key/foreign key (PK/FK) joins. Finally, if the mapping function $h$ performs some form of semantic value mapping via embeddings or a thesaurus (e.g., $h($"Big Apple"$) \rightarrow$ "New York"), the joinability is  \textit{semantic}. In practice, we are interested in all three joinability cases: the exact-, PK/FK-, and semantic-joinability.

\section{\method{}: Generating Corpora Using Ontologies and LLMs}
\label{sec:tables}
\vspace{-3mm}

In this section, we describe \method, which generates a domain-specific benchmark from semantic types and relationships extracted from an ontology. \method consists of two modules: base table generation, described in this section, and table perturbations that we describe in \Cref{sec:table_perturb}.

\subsection{Ontologies as Sources of Domain-specific Semantic Data Types \& Joinability}
\label{sec:tables:input}
\vspace{-2mm}

\begin{figure*}[t]
    \centering
    \includegraphics[width=.99\textwidth]{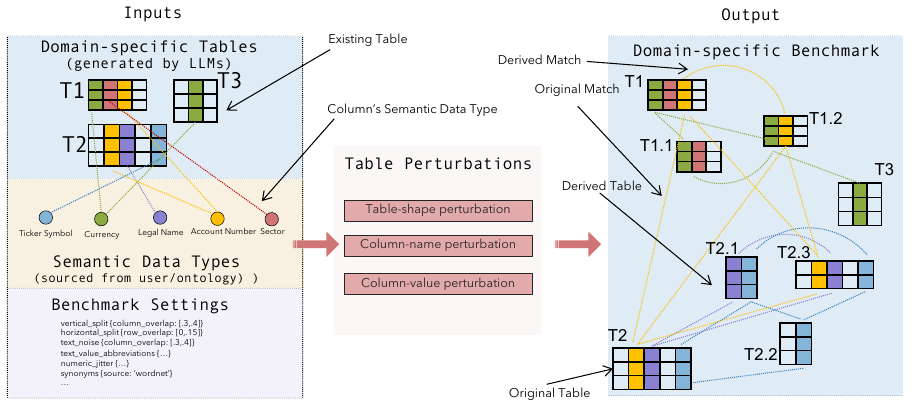}
    \vspace{-4mm}
    \caption{\myTag{\method details:} The process of using a set of base tables and their associated semantic data types, in order to generate a set of derived tables and ground truth for dataset discovery tasks such as joinability.}
    \label{fig:table-to-benchmark}
    % \vspace{-3mm}
\end{figure*}

During the last two decades, various domain-specific ontologies have been developed. Those include FIBO for finance~\cite{fibo}, the Gene Ontology \cite{gene2004gene} for genomics, SNOMED CT \cite{donnelly2006snomed} and Dron \cite{hanna2013building} for healthcare, Amazon's product knowledge graph \cite{dong2018challenges} for retail, and many others. We argue that these ontologies contain rich semantic data types and relationships across different ontology classes/entities that we can leverage to generate complex datasets for data discovery tasks. However, to this date, these semantic data types and relationships have not been widely used for evaluating dataset discovery methods, mainly because real-world databases that adopt these ontologies are siloed behind corporate and governmental regulations.

\para{Ontology to base tables generation} 
\method{}, uses the domain-specific ontology to drive the \textit{table schema} generation. A domain ontology, such the one depicted on the left of \Cref{fig:ontology-to-table}, provides such a rich and expressive data model to capture a variety of real-world relationships between entities. We adopt OWL~\cite{mcguinness2004owl} for domain ontologies, where real-world entities are captured as \textit{concepts}, each concept has zero or more \textit{data properties}, describing the concept, and zero or more \textit{object properties}, capturing its \textit{relationships} with other concepts.

In \method{}, we map a concept into a table and its data properties are converted to the columns in the table. For example, in \Cref{fig:ontology-to-table}, the concept ``Postal Address'' has data properties such as ``Address Line 1'', ``Address Line 2'', ``City'', ``State'', ``Zipcode'', etc. This information can naturally be mapped to the schema of table ``Postal Address'' as seen in the middle of the same figure. If a concept does not contain multiple data properties, we can group several concepts into one table, with each associated data property becoming a column.

\para{Extracting joinable relationships from an ontology} 
Once we have generated the table schema based on the concepts and their data properties from the ontology, we can also obtain \textit{joinability relationships} between tables from the object properties in the ontology. We follow the domain and range of an object property to connect two tables. For example, in \Cref{fig:ontology-to-table}, the concept ``Organization'' has a relation ``lists'' which connects to another concept ``Listed Security''. Therefore, the table ``Listed Security'' should include a column called ``listedBy'' that can be joined with a column in the table ``Organization''. Here we then annotate which column in table ``Organization'' is related to ``Listed Security''. In this case, the ``listedBy'' column points to the primary key column ``Organization.id'' of the table ``Organization''. 
% Considering the number of base tables and column relations are not large (usually less than 100), this process minimal human annotation effort to prepare the inputs for our pipeline. 

\subsection{LLMs for Table-value Generation}
\label{sec:base_tables_generation}
\vspace{-2mm}

Traditional data generators such as Faker\footnote{\url{https://github.com/joke2k/faker}} generate the contents of each column independently from the rest of the table, with a user-defined generation function. Approaches like Faker, however, only cover common semantic types and cannot generate suitable contents for domain-specific ones. Moreover, Faker-like approaches do not maintain the value dependencies between rows, i.e., the columns ``Zipcode'' and ``City'' should match. In the case of Faker, one would have to use a user-defined dictionary of mappings between zip codes and cities.

\para{Why LLMs} LLMs can address the above issues since they are good at generating domain-specific contents since the models are trained on a vast amount of data covering diverse domains and semantic types. Consequently, this makes them highly suitable for generating high-quality and relevant content for specific domains. In addition, LLMs are good with using \textit{context} to generate factual data. For example, we observe that a LLM, \texttt{text-davinci-003}, generates nearly perfect matches between cities and zip codes in our preliminary experiments, without further actions required by a user. The contents of the table in the bottom right of \Cref{fig:ontology-to-table} was generated from values extracted with a simple prompt from the \texttt{text-davinci-003} LLM. To the best of our knowledge, such a high-quality dataset is not possible to be generated with existing synthetic data generation pipelines. To maintain the dependency between the row entries, the table contents are generated row by row.

\para{Joinability relationships} In \method{}, the semantic type relations are naturally converted to the relations between columns with the following definition. For this paper, we mainly consider the column dependency relations, where column $i$ from table a $R$ depends on column $j$ from table $S$ if $R_i \subseteq S_j$. We now exemplify how to satisfy the constraints of input column relations, i.e., how the tables should are generated according to the inter-column dependencies like Primary Key to Foreign Key relations. 
Namely, if the values of column $R_i$ comes from column $S_j$ (i.e., $S_j$ is the primary key and $R_i$ is the foreign key), after generating table $R$, we should make sure the column $R_i$ includes all the entries of column $S_j$ when generating table $S$. To achieve this goal, while generating the rows of table $S$, we parse the entries in column $R_i$ to the prompt and require column $S_j$ should include these entries. In the prompt see in \Cref{lst:prompt}, we show how to generate table $S$ conditional on column $R_i$.

For the case where multiple relations between multiple columns and tables exist, we use Algorithm~\ref{alg:algo1}. First, we build a graph $G$ between tables, where table $A$ has an edge pointing to table $B$ if table $A$ depends on table $B$ (i.e., there exists a column $i$ from table $A$ comes from column $j$ in table $B$). Then, we generate the tables one-by-one along the graph edges. Obviously, we should first generate tables with $in\_degree = 0$ since these tables can be generated without any column-wise pre-conditions. The generated tables are added to a set $Q$. Finally, given $Q$, we generate the table $T$ whose column-wise dependencies only come from the tables in the set $Q$ (the predecessors of $T$ come from $Q$) in the next round.

\vspace{-2mm}
\section{Proposed LLM-based Table Perturbations}
\label{sec:table_perturb}
\vspace{-3mm}

The column relations between the base tables offer natural exact-joinable relationships. However, more diverse semantically-joinable relationships are needed  for evaluating data discovery algorithms. To this end, the joinable column pairs in the base tables need to provide semantically consistent column contents, that may not necessarily be identical. For instance, a column ``country'' can contain values such as ``US'' or ``USA'' while another, semantically-joinable column may contain the term ``United States''. To address this,  we employ LLMs to perturb the contents of the joinable column pairs, generating semantically equivalent but not identical column contents. Moreover, we can also create synthetic table pairs with ground truth through table perturbations (see table~\ref{tab:perturb_funcs}), which expands the dataset and enriches the inter-table relationships.

\subsection{LLM-based Semantically Joinable Relationship Generation}
\vspace{-2mm}

Discovering semantically joinable column pairs is a challenging task because these pairs do not require the column contents to be identical. Instead, if the entries of two columns are semantically equivalent, the column pair is considered semantically joinable. Following the approach above to generate base tables, a group of exact-joinable column pairs can be obtained. To create more challenging semantically-joinable column pairs, we can perturb exact-joinable columns' contents with semantically equivalent entries.

\para{LLMs for semantically-safe value perturbations}
A naive approach to perturb a column is to add noise to its entries \cite{koutras2021valentine,lee2007etuner}, such as introducing typos or replacing them with abbreviations or synonyms. However, these approaches cannot guarantee the perturbed entries still keep their original semantic meanings. In addition, naively adding typos to columns such as ``URL'' or ``Account Number'' could change the original semantics. Similarly, replacing ``Address'' with synonyms could lead to a different address, as ``123 Main Street'' is not equivalent to ``123 Principal Street'' or ``123 Main Boulevard''. Finally, adding noise should take the context of a table schema into consideration to generate more diverse semantically equivalent variations of the column's entries. For example, ``CAD'' in a column ``Currency'' cannot be safely expanded to ``Canadian Dollar'' using traditional approaches. We observed that LLMs do address the aforementioned issues and generate accurate and diverse variations of the original entries. For example, given a street name ``123 Main Street'' the LLM is able to abbreviate it to ``123 Main St.'' i.e., without breaking the semantics of its value.

\subsection{Table-level Perturbations}
\vspace{-2mm}

\para{Table-shape perturbations}
In addition to leveraging a LLM to generate diverse joinable table relations, we can also fabricate synthetic joinable relations through table-shape perturbations (see Table~\ref{tab:perturb_funcs}). By vertically splitting the base tables \cite{koutras2021valentine}, we create sub-tables with joinable columns that share at least one common column. As see in \Cref{fig:table-to-benchmark}, \method{} adopts this method in order to split tables while generating additional tables and joins among them.

\para{Column-value perturbations} To introduce additional diversity to the generated benchmarks, we  use column-value and schema-level perturbations (\Cref{tab:perturb_funcs}). These perturbations can  pose additional challenges to the downstream data discovery algorithms. For example, our schema perturbation could rewrite the column headers with their cryptic names, i.e. ``Person ID'' can be encrypted to ``PID''. When the data discovery algorithm (such as COMA~\cite{do2002coma}) leverages the information of column headers for column relation discovery, the cryptic column headers will significantly reduce the token-level and semantic-level similarity between joinable columns, making it more difficult to differentiate the joinable pairs.

\vspace{-3mm}
\section{Experimental Evaluation}
\label{sec:experiments}
\vspace{-3mm}

In this section, we mainly focus on two data discovery tasks: $i)$ exact-joinability and $ii)$ semantic joinability detection  as defined in \Cref{sec:problem}.

% \subsection{Experimental Setup}
% \label{sec:exp:setup}

\para{Datasets} We used \method{} to generate  datasets for three different domains, including finance, pharma, and healthcare domains. To generate the finance and pharma domain datasets, we start with the semantic types and relations from the financial industry business ontology (FIBO)~\cite{fibo} and DrugBank~\cite{wishart2018drugbank} respectively. For the healthcare domain, we extract part of the MIMIC-III dataset as the base tables~\cite{johnson2016mimic, johnson2019mimic} and used perturbations only, since the base tables are already provided in the original dataset. For the finance and pharma domains, we generated the 20 base tables of maximum $1,000$ rows. The healthcare base tables come from a sample of MIMIC-III dataset (sampled at most $2,000$ rows from each table), including 32 base tables. Table~\ref{tab:data_summary} describes the generated datasets. 
% With the base tables, we leverage the table perturbations in sec~\ref{sec:table_perturb} to expand the base tables into 100 tables in finance and pharma domain, and 150 tables in healthcare domain. We also use LLM to rewrite the column cells with semantically equivalent representations. 

\para{Discovery methods} The data discovery methods are initially designed for schema matching tasks but they can also be used for the joinable relationship detection tasks where we can match the same dataset. In this experiment, we included four methods, namely: $i)$ COMA \cite{do2002coma}, $ii)$ Jaccard-Levenshiten (JL) \cite{koutras2021valentine}, $iii)$ Similarity Flooding (SF) \cite{melnik2002similarity} and $iv)$ Starmie \cite{fan2022semantics}, a recent discovery approach that leverages column representations using a pre-trained language model.

\para{Metrics}
We evaluated the performance of these methods using precision/recall and top-$k$ precision. While evaluating the precision/recall, column pairs with prediction score $>0.5$ are treated as a positive predictions. The top-$k$ precision is defined as: top-$k$ precision = (\# of true positives in top $k$ predictions)/$k$. The top-$k$ predictions refer to the column pairs with largest prediction scores. In our experiments, we used $k=1,3,5$. 

% \begin{equation}
% \small
%     \text{top-$k$ precision} = \frac{\text{\# of true positives in top $k$ predictions}}{k} \nonumber
% \end{equation}

\begin{table}[t]
\resizebox{\textwidth}{!}{
\begin{tabular}{|c|c|c|c|c|c|c|}
\hline
\textbf{Dataset}    & \textbf{\# of base tables} & \textbf{\# of tables} & \textbf{avg \# of rows} & \textbf{avg \# of columns} & \textbf{\# of exact joins} & \textbf{\# of semantic joins} \\ \hline
Finance & 20    & 100   & 604.1 & 4.62  & 301   & 540 \\ \hline
Pharma  & 20    & 100   & 129.7 & 4.97  & 905   & 1646\\ \hline
Healthcare  & 32    & 150   & 1137.4    & 6.22  & 896   & 6613\\ \hline
\end{tabular}}
\vspace{.5mm}
\caption{Statistics of three generated datasets.\vspace{-5mm}}
\label{tab:data_summary}
\end{table}

% \huzefa{Do you all want to bold the best or winning results in each of the tables?}

\vspace{-2mm}
\subsection{Results}
\vspace{-2mm}

\para{Performance of dataset discovery methods}
We evaluated the performance of 4 methods on the exact-joinable and semantically-joinable relationship detection tasks. In general, for different domains, these methods show significantly different performance, highlighting the need for methods like \method{}, in order to generate diverse datasets. 
Tables~\ref{tab:prec_recall} and~\ref{tab:top_K_prec} confirm that semantically joinable relation discovery is hard in most cases, achieving lower $F_1$ scores (dropping up to $10\%$). Among all these methods, traditional approaches such as COMA and Jaccard Levenshiten significantly outperform the other two. Both COMA and Jaccard Levenshiten methods mainly leverage the column cell overlaps and schema similarity to determine the joinable column pairs, which discover most of the joinable relations. In contrast, Starmie relies on column embeddings learned from a language model. Our experiments show that the similarity between embeddings is not a reliable signal for joinable relationship detection. This reveals the importance of integrating current language model-based column representations with traditional column-value/schema containment signals.

\begin{table}[t]
\resizebox{\textwidth}{!}{
\begin{tabular}{|c|ccc|ccc|}
\hline
                 & \multicolumn{3}{c|}{\textbf{Exact Joinable Relations}}                                                                & \multicolumn{3}{c|}{\textbf{Semantically Joinable Relations}}                                                                  \\ \hline
\textbf{Methods} & \multicolumn{1}{c|}{Finance}                    & \multicolumn{1}{c|}{Pharma}                     & Healthcare        & \multicolumn{1}{c|}{Finance}                    & \multicolumn{1}{c|}{Pharma}                     & Healthcare                 \\ \hline
COMA             & \multicolumn{1}{c|}{0.68 (0.55, 0.87)}          & \multicolumn{1}{c|}{\textbf{0.85 (0.77, 0.96)}} & 0.68 (0.52, 0.99) & \multicolumn{1}{c|}{0.64 (0.59, 0.69)}          & \multicolumn{1}{c|}{\textbf{0.85 (0.77, 0.95)}} & \textbf{0.75 (0.60, 0.98)} \\ \hline
JL               & \multicolumn{1}{c|}{\textbf{0.74 (0.81, 0.67)}} & \multicolumn{1}{c|}{0.39 (0.87, 0.25)}          & 0.32 (0.25, 0.43) & \multicolumn{1}{c|}{\textbf{0.64 (0.84, 0.52)}} & \multicolumn{1}{c|}{0.38 (0.87, 0.24)}          & 0.30 (0.26, 0.38)          \\ \hline
SF               & \multicolumn{1}{c|}{0.10 (0.38, 0.06)}          & \multicolumn{1}{c|}{0.22 (0.91, 0.13)}          & 0.11 (0.52, 0.06) & \multicolumn{1}{c|}{0.08 (0.42, 0.05)}          & \multicolumn{1}{c|}{0.23 (0.93, 0.13)}          & 0.11 (0.60, 0.06)          \\ \hline
Starmie          & \multicolumn{1}{c|}{0.12 (0.08, 0.30)}          & \multicolumn{1}{c|}{0.13 (0.10, 0.17)}          & 0.06 (0.28, 0.03) & \multicolumn{1}{c|}{0.13 (0.09, 0.26)}          & \multicolumn{1}{c|}{0.13 (0.11, 0.17)}          & 0.06 (0.32, 0.03)          \\ \hline
\end{tabular}}
\vspace{.5mm}
\caption{\textbf{Performance evaluation:} Four methods performance over the 3 different domains-specific datasets. For each method, the prediction score threshold is set at $0.5$. In each cell, the performance is shown in the form of $F_1$ score (precision, recall).\vspace{-5mm}}
\label{tab:prec_recall}
\end{table}

\para{Discovering semantically joinable columns}
Semantically joinable relation detection is challenging when two columns have different column headers/values that are semantically equivalent. In our experience, such semantically joinable pairs are commonplace in real-world data lakes. \method{} provides a significant advantage over the traditional split-based approaches in generating such challenging semantically joinable column pairs. Since the exact-joinable pairs have covered the cases for columns with overlapping instances, here we focus on analyzing the non-exactly but semantically joinable pairs. Specifically, we divide semantically joinable pairs into two categories, ``difficult'' and ``easy'' pairs, referring to the column pairs with different or identical column headers, respectively.

Table \ref{tab:diff_pairs} shows that all four methods fail to discover the difficult pairs, in particular Starmie, which uses a pre-trained language model to encode columns. However, the learned column embeddings cannot capture the semantic similarity between the difficult column pair cell values. Therefore, our experiments demonstrate the necessity to improve the quality of column representations in order to discover the semantic mappings between column cells. As a comparison, COMA, leveraging both schema and instance information, does better in predicting the easy joinable pairs. On the other hand, the other methods do not adequately leverage the schema information and fail to detect the easy pairs.

\begin{table}[]
\resizebox{\textwidth}{!}{
\begin{tabular}{|c|ccc|ccc|}
\hline
                 & \multicolumn{3}{c|}{\textbf{Exact Joinable Relations}}                                                                      & \multicolumn{3}{c|}{\textbf{Semantically Joinable Relations}}                                                               \\ \hline
\textbf{Methods} & \multicolumn{1}{c|}{Finance}                   & \multicolumn{1}{c|}{Pharma}                    & Healthcare                & \multicolumn{1}{c|}{Finance}                   & \multicolumn{1}{c|}{Pharma}                    & Healthcare                \\ \hline
COMA             & \multicolumn{1}{c|}{0.53, 0.40, 0.29}          & \multicolumn{1}{c|}{\textbf{0.58, 0.50, 0.39}} & \textbf{0.89, 0.78, 0.64} & \multicolumn{1}{c|}{0.52, 0.41, 0.31}          & \multicolumn{1}{c|}{\textbf{0.58, 0.51, 0.41}} & \textbf{0.91, 0.81, 0.67} \\ \hline
JL               & \multicolumn{1}{c|}{\textbf{0.52, 0.41, 0.31}} & \multicolumn{1}{c|}{0.55, 0.48, 0.39}          & 0.57, 0.48, 0.38          & \multicolumn{1}{c|}{\textbf{0.52, 0.42, 0.32}} & \multicolumn{1}{c|}{0.55, 0.48, 0.39}          & 0.51, 0.43, 0.35          \\ \hline
SF               & \multicolumn{1}{c|}{0.36, 0.24, 0.17}          & \multicolumn{1}{c|}{0.47, 0.35, 0.30}          & 0.53, 0.42, 0.37          & \multicolumn{1}{c|}{0.37, 0.25, 0.18}          & \multicolumn{1}{c|}{0.49, 0.37, 0.31}          & 0.6, 0.45, 0.40           \\ \hline
Starmie          & \multicolumn{1}{c|}{0.07, 0.06, 0.05}          & \multicolumn{1}{c|}{0.13, 0.11, 0.10}          & 0.42, 0.30, 0.25          & \multicolumn{1}{c|}{0.08, 0.07, 0.06}          & \multicolumn{1}{c|}{0.14, 0.11, 0.10}          & 0.42, 0.31, 0.26          \\ \hline
\end{tabular}}
\vspace{1mm}
\caption{\textbf{Performance evaluation:} Top-$k$ accuracy of baselines over the 3 different domains-specific datasets. In each cell, the performance is shown in the form of top $1,3,5$ precision. \vspace{-5mm}}
\label{tab:top_K_prec}
\end{table}

\begin{table}[]
\resizebox{\textwidth}{!}{
\begin{tabular}{|c|c|ccccc|ccccc|}
\hline
& & \multicolumn{5}{c|}{\textbf{Difficult Non-exactly Joinable Pairs}}   & \multicolumn{5}{c|}{\textbf{Easy Non-exactly Joinable Pairs}} \\ \hline
\textbf{Dataset} & \begin{tabular}[c]{@{}c@{}}Non-exact join pairs/\\ All semantic join pairs\end{tabular} & \multicolumn{1}{c|}{$\#$ pairs} & \multicolumn{1}{c|}{Coma}   & \multicolumn{1}{c|}{JL} & \multicolumn{1}{c|}{SF} & Starmie & \multicolumn{1}{c|}{$\#$ pairs} & \multicolumn{1}{c|}{Coma}    & \multicolumn{1}{c|}{JL} & \multicolumn{1}{c|}{SF} & Starmie \\ \hline
Finance    & 180/540                                                                             & \multicolumn{1}{c|}{134}         & \multicolumn{1}{c|}{24/134} & \multicolumn{1}{c|}{9/134}   & \multicolumn{1}{c|}{2/134}   & 18/134  & \multicolumn{1}{c|}{46}         & \multicolumn{1}{c|}{46/46}   & \multicolumn{1}{c|}{3/46}    & \multicolumn{1}{c|}{2/46}    & 6/46    \\ \hline
Pharma       & 50/1646                                                                             & \multicolumn{1}{c|}{27}          & \multicolumn{1}{c|}{0/27}   & \multicolumn{1}{c|}{0/27}    & \multicolumn{1}{c|}{1/27}    & 4/27    & \multicolumn{1}{c|}{23}         & \multicolumn{1}{c|}{23/23}   & \multicolumn{1}{c|}{0/23}    & \multicolumn{1}{c|}{3/23}    & 4/23    \\ \hline
Healthcare & 890/6613                                                                            & \multicolumn{1}{c|}{70}          & \multicolumn{1}{c|}{26/70}  & \multicolumn{1}{c|}{0/70}    & \multicolumn{1}{c|}{0/70}    & 0/70    & \multicolumn{1}{c|}{820}        & \multicolumn{1}{c|}{820/820} & \multicolumn{1}{c|}{35/820}  & \multicolumn{1}{c|}{56/820}  & 30/820  \\ \hline
\end{tabular}}
\vspace{1mm}
\caption{\textbf{Performance evaluation:} Four methods over the semantically joinable but not exactly joinable pairs. We decompose these pairs into two categories: 1) difficult non-exactly joinable pairs where two column headers as well as cells are different and 2) easy non-exactly joinable pairs where two column headers are identical. The table shows the number of correct predictions made by different baselines.}
\label{tab:diff_pairs}
\end{table}

\para{Base table evaluation}
We also examined the performance of four methods using newly generated base tables in three domains. As a comparison, all methods are also evaluated over 3 widely used data discovery corpora, including ``Gosales'',  ``TPC-H'' and ``TPC-DS''. Here, we choose the semantically joinable relation detection task for the evaluation. 
Tables~\ref{tab:base_tables} suggests that \method generated base tables are much more challenging than the existing corpora, as all methods perform much worse on our base tables. As we motivated in \Cref{sec:intro}, existing corpora lack enough schema and cell value complexity, and the joinable column pairs have similar column headers and cell-values. Our experiments demonstrate that without a realistic domain-specific dataset, the performance of data discovery methods cannot be truly stress-tested.

\begin{table}[]
\resizebox{\textwidth}{!}{
\begin{tabular}{|c|ccc|ccc|}
\hline
& \multicolumn{3}{c|}{\textbf{$F_1$ (Precision, Recall)}} & \multicolumn{3}{c|}{\textbf{Top-$k$ Precision ($k$ = 1,3,5)}} \\\hline
\textbf{Methods} & \multicolumn{1}{c|}{Finance} & \multicolumn{1}{c|}{Pharma} & Healthcare    & \multicolumn{1}{c|}{Finance}    & \multicolumn{1}{c|}{Pharma} & Healthcare \\ \hline
Coma    & \multicolumn{1}{c|}{0.23 (0.19, 0.29)} & \multicolumn{1}{c|}{0.65 (0.54, 0.82)} & 0.61 (0.44, 0.99)  & \multicolumn{1}{c|}{0.41, 0.26, 0.20} & \multicolumn{1}{c|}{0.63, 0.59, 0.58} & 0.94, 0.83, 0.78 \\ \hline
JL & \multicolumn{1}{c|}{0.13 (0.33, 0.08)} & \multicolumn{1}{c|}{0.32 (0.64, 0.21)} & 0.35 (0.24, 0.67)  & \multicolumn{1}{c|}{0.59, 0.42, 0.29} & \multicolumn{1}{c|}{0.89, 0.75, 0.68} & 0.32, 0.30, 0.28 \\ \hline
SF & \multicolumn{1}{c|}{0.0 (0.0, 0.0)}    & \multicolumn{1}{c|}{0.0 (0.0, 0.0)}    & 0.002 (0.5, 0.001) & \multicolumn{1}{c|}{0.22, 0.15, 0.10} & \multicolumn{1}{c|}{0.71, 0.59, 0.57} & 0.56, 0.47, 0.37 \\ \hline
Starmie & \multicolumn{1}{c|}{0.05 (0.02, 0.42)} & \multicolumn{1}{c|}{0.06 (0.04, 0.26)} & 0.04 (0.15, 0.02)  & \multicolumn{1}{c|}{0.26, 0.10, 0.11} & \multicolumn{1}{c|}{0.13, 0.10, 0.10} & 0.52, 0.43, 0.35 \\ \hline
& \multicolumn{1}{c|}{Gosales} & \multicolumn{1}{c|}{TPC-H} & TPC-DS & \multicolumn{1}{c|}{Gosales}    & \multicolumn{1}{c|}{TPC-H} & TPC-DS \\ \hline
Coma    & \multicolumn{1}{c|}{1.00 (1.00, 1.00)} & \multicolumn{1}{c|}{0.86 (0.75, 1.00)} & 0.45 (0.81, 0.31)  & \multicolumn{1}{c|}{1.00, 0.87, 0.62} & \multicolumn{1}{c|}{1.00, 0.83, 0.58} & 0.35, 0.31, 0.24 \\ \hline
JL & \multicolumn{1}{c|}{0.80 (1.00, 0.60)} & \multicolumn{1}{c|}{0.72 (1.00, 0.56)} & 0.37 (1, 0.23)  & \multicolumn{1}{c|}{1.00, 0.87, 0.72} & \multicolumn{1}{c|}{1.00, 0.82, 0.69} & 0.34, 0.29, 0.23 \\ \hline
SF & \multicolumn{1}{c|}{0.0 (0.0, 0.0)}    & \multicolumn{1}{c|}{0.0 (0.0, 0.0)}    & 0.00 (0.00, 0.00) & \multicolumn{1}{c|}{1.00, 0.83, 0.61} & \multicolumn{1}{c|}{1.00, 0.79, 0.57} & 0.34, 0.25, 0.17 \\ \hline
Starmie & \multicolumn{1}{c|}{1.00 (1.00, 1.00)} & \multicolumn{1}{c|}{0.83 (0.77, 0.91)} & 0.84 (0.98, 0.74)  & \multicolumn{1}{c|}{0.83, 0.67, 0.55} & \multicolumn{1}{c|}{0.82, 0.61, 0.52} & 0.46, 0.33, 0.15 \\ \hline
\end{tabular}}
\vspace{1mm}
\caption{\textbf{Performance evaluation:} Four methods' performance on the base tables and other public datasets (Gosalse, TPC-H, TPC-DS). The performance is evaluated over the semantically joinable relation discovery task.\vspace{-4mm}}
\label{tab:base_tables}
\end{table}

\vspace{-2mm}
\section{Related Work}
\label{sec:relatedwork}
\vspace{-2mm}

Since we have addressed existing corpora in the introduction and \Cref{tab:related-works}, we refrain from re-iterating over those in this section. Instead, we turn to synthetic tabular data generation methods. 

While the generation of images \cite{karras2020a} and text \cite{subramanian2017a} has been thoroughly studied, the generation of synthetic tabular data has received less attention in recent machine learning literature. 
Recent approaches to generating tabular data employ generative adversarial networks (\cite{choi2017a,park2018a,mottini2018a,xu2019a, koivu2020a}) or variational autoencoders (\cite{xu2019a,ma2020a,vardhan2020a, darabi2021a}). The current state-of-the-art approach, CTGAN \cite{xu2019a}, emphasizes the conditional distributions among the features to generate semantically meaningful data.
Transformer-based models have also been developed for tabular data classification \cite{arik2021tabnet,somepalli2021a,kossen2021a} and for learning joint representations of tabular and textual data (\cite{yin2020a}). \cite{padhi2021a} investigated the generation of multi-variate time series data using a transformer architecture based on BERT \cite{devlin2018a}.

In contrast to these works, \method{} focuses on the orthogonal problem of generating domain-specific corpora with joinable relationships, in order to evaluate dataset discovery methods. In our prototype, we have used a general LLM (\texttt{text-davinci-003}), using prompts to generate values for tabular data. However, our work can benefit from other general data generation frameworks like CTGAN~\cite{xu2019a} as well as GReaT~\cite{borisov2022language} in order to populate the tables schemas extracted from ontologies. Finally, in the future our approach can adopt domain-specific GAN-based generation approaches for e.g., health records \cite{baowaly2019synthesizing,koivu2020a,choi2017a}, or passenger records \cite{mottini2018a}, etc.

% \vspace{-2mm}
\section{Conclusion}
\label{sec:conclusion}
% \vspace{-3mm}

In this paper, we presented \method, the first corpus generator that leverages $i)$ ontologies to extract domain-specific data types in order to create tabular dataset schemas which are then $ii)$ populated using LLMs. \method{} requires minimal user intervention in order to generate large complex schemata of datasets, including joinability relationships. We have used \method{} to generate corpora from three different domains, namely financial, pharma and healthcare, and we have shown that such domain-specific datasets create challenges not found in existing public corpora.

%\vspace{-2mm}
\section*{Customer Problem Statement}
% \vspace{-3mm}
%Authors are required to include a statement of the customer problem that is being solved in their work. Authors should discuss both positive and negative outcomes, if any. For instance, authors should discuss a) which customer problem may benefit from this research,

% \xiao{All papers must include a minimum of 1-paragraph that discusses the Amazon customer problem of the topic presented and must state whether this is a \bf short-term, long-term or existing impact. }

\para{Specializing models for clients} 
When we onboard new customers from various domains on Amazon DataZone\footnote{\url{https://aws.amazon.com/datazone/}}, we need to deploy a highly effective dataset discovery model. In our experience, some dataset discovery methods may do well with finance databases, but not for chemical databases, etc. Ideally, the data provided by one customer should be used for training dataset discovery and matching methods, before being shipped to customers. However, due to privacy and regulatory reasons, access to real customer data can be difficult and limited. In addition, even if we had access to such customer data, we would be missing ground truth to evaluate dataset discovery methods. 
Currently, we are leveraging \method to generate domain-specific datasets alongside matches, in order to train internal ML-based data discovery methods, and to establish hyper-parameters before those methods are deployed in production for both internal or external clients.
%b) which customers may be put at disadvantage from this research, 

\para{Generating corpora for training Bedrock/M* models} 
Modern cloud services are increasingly reliant on data-driven technologies such as machine learning and data mining. The development and operation of such services require large amounts of high quality data to harness the power of artificial intelligence to drive innovation, improve decision-making and create value for businesses and customers. Obtaining diverse, high quality and high volume datasets for developing solutions for customers from a wide range of domains can be extremely challenging due to reasons such as data quality issues, data privacy concerns, data access and ownership, data bias and data scalability. We argue that our method \method can be used to generate large-scale realistic datasets for table representation learning. Those tasks may include table summarization, semantic data types detection on columns, knowledge graph construction, etc.

% %c) what are the consequences of failure of the system, and 
% If the generated dataset does not adhere to the semantic data types of the target customer, we run the risk of deploying an ineffective model -- a generic model would have been better in that case.
% % d) whether the task/method leverages biases in the data.
% Finally, we are not aware of biases in the data that is generated, that may put Amazon at risk.

% % {\bf All submissions must include a Customer Problem Statement in order to be eligible for review. The Customer Problem Statement must be included in your 8-page limit.}

\bibliographystyle{unsrt}
\bibliography{references}

\begin{thebibliography}{10}

\bibitem{doan2012principles}
AnHai Doan, Alon Halevy, and Zachary Ives.
\newblock {\em Principles of data integration}.
\newblock Elsevier, 2012.

\bibitem{ritchie2015methods}
Marylyn~D Ritchie, Emily~R Holzinger, Ruowang Li, Sarah~A Pendergrass, and Dokyoon Kim.
\newblock Methods of integrating data to uncover genotype--phenotype interactions.
\newblock {\em Nature Reviews Genetics}, 16(2):85--97, 2015.

\bibitem{dong2021efficient}
Yuyang Dong, Kunihiro Takeoka, Chuan Xiao, and Masafumi Oyamada.
\newblock Efficient joinable table discovery in data lakes: A high-dimensional similarity-based approach.
\newblock In {\em 2021 IEEE 37th International Conference on Data Engineering (ICDE)}, pages 456--467. IEEE, 2021.

\bibitem{fan2022semantics}
Grace Fan, Jin Wang, Yuliang Li, Dan Zhang, and Ren{\'e}e Miller.
\newblock Semantics-aware dataset discovery from data lakes with contextualized column-based representation learning.
\newblock {\em PVLDB}, 2023.

\bibitem{lee2007etuner}
Yoonkyong Lee, Mayssam Sayyadian, AnHai Doan, and Arnon~S. Rosenthal.
\newblock {ETuner:} tuning schema matching software using synthetic scenarios.
\newblock {\em VLDBJ}, 16(1):97–122, 2007.

\bibitem{koutras2021valentine}
Christos Koutras, George Siachamis, Andra Ionescu, Kyriakos Psarakis, Jerry Brons, Marios Fragkoulis, Christoph Lofi, Angela Bonifati, and Asterios Katsifodimos.
\newblock Valentine: Evaluating matching techniques for dataset discovery.
\newblock In {\em 2021 IEEE 37th International Conference on Data Engineering (ICDE)}, pages 468--479. IEEE, 2021.

\bibitem{nargesian2018table}
Fatemeh Nargesian, Erkang Zhu, Ken~Q Pu, and Ren{\'e}e~J Miller.
\newblock Table union search on open data.
\newblock In {\em VLDB}, 2018.

\bibitem{hulsebos2021gittables}
Madelon Hulsebos, Cagatay Demiralp, and Paul Groth.
\newblock Gittables: A large-scale corpus of relational tables.
\newblock {\em arXiv preprint arXiv:2106.07258}, 2021.

\bibitem{tpcdi}
Meikel Poess, Tilmann Rabl, Hans-Arno Jacobsen, and Brian Caufield.
\newblock {TPC-DI}: The first industry benchmark for data integration.
\newblock In {\em VLDB}, 2014.

\bibitem{gaulton2012chembl}
Anna Gaulton, Louisa~J Bellis, A~Patricia Bento, Jon Chambers, Mark Davies, Anne Hersey, Yvonne Light, Shaun McGlinchey, David Michalovich, Bissan Al-Lazikani, et~al.
\newblock Chembl: a large-scale bioactivity database for drug discovery.
\newblock {\em Nucleic acids research}, 40(D1):D1100--D1107, 2012.

\bibitem{magellandata}
Sanjib Das, AnHai Doan, Paul~Suganthan G.~C., Chaitanya Gokhale, Pradap Konda, Yash Govind, and Derek Paulsen.
\newblock The magellan data repository.
\newblock \url{https://sites.google.com/site/anhaidgroup/useful-stuff/data}.

\bibitem{donnelly2006snomed}
Kevin Donnelly et~al.
\newblock Snomed-ct: The advanced terminology and coding system for ehealth.
\newblock {\em Studies in health technology and informatics}, 121:279, 2006.

\bibitem{nambiar2006making}
Raghunath~Othayoth Nambiar and Meikel Poess.
\newblock The making of tpc-ds.
\newblock In {\em VLDB}, volume~6, pages 1049--1058, 2006.

\bibitem{DBLP:journals/ijon/HernandezEACR22}
Mikel Hernandez, Gorka Epelde, Ane Alberdi, Rodrigo Cilla, and Debbie Rankin.
\newblock Synthetic data generation for tabular health records: {A} systematic review.
\newblock {\em Neurocomputing}, 493:28--45, 2022.

\bibitem{borisov2023language}
Vadim Borisov, Kathrin Seßler, Tobias Leemann, Martin Pawelczyk, and Gjergji Kasneci.
\newblock Language models are realistic tabular data generators, 2023.

\bibitem{DBLP:conf/emnlp/PetroniRRLBWM19}
Fabio Petroni, Tim Rockt{\"{a}}schel, Sebastian Riedel, Patrick S.~H. Lewis, Anton Bakhtin, Yuxiang Wu, and Alexander~H. Miller.
\newblock Language models as knowledge bases?
\newblock In {\em {EMNLP-IJCNLP}}, pages 2463--2473, 2019.

\bibitem{fibo}
Mike Bennett.
\newblock The financial industry business ontology: Best practice for big data.
\newblock {\em Journal of Banking Regulation}, 14(3-4):255--268, 2013.

\bibitem{rahm2001survey}
Erhard Rahm and Philip~A Bernstein.
\newblock A survey of approaches to automatic schema matching.
\newblock {\em VLDBJ}, 10(4):334--350, 2001.

\bibitem{gene2004gene}
Gene~Ontology Consortium.
\newblock The gene ontology (go) database and informatics resource.
\newblock {\em Nucleic acids research}, 32(suppl\_1):D258--D261, 2004.

\bibitem{hanna2013building}
Josh Hanna, Eric Joseph, Mathias Brochhausen, and William~R Hogan.
\newblock Building a drug ontology based on rxnorm and other sources.
\newblock {\em Journal of biomedical semantics}, 4:1--9, 2013.

\bibitem{dong2018challenges}
Xin~Luna Dong.
\newblock Challenges and innovations in building a product knowledge graph.
\newblock In {\em Proceedings of the 24th ACM SIGKDD International conference on knowledge discovery \& data mining}, pages 2869--2869, 2018.

\bibitem{mcguinness2004owl}
Deborah~L McGuinness, Frank Van~Harmelen, et~al.
\newblock Owl web ontology language overview.
\newblock {\em W3C recommendation}, 10(10):2004, 2004.

\bibitem{do2002coma}
Hong-Hai Do and Erhard Rahm.
\newblock Coma—a system for flexible combination of schema matching approaches.
\newblock In {\em VLDB'02: Proceedings of the 28th International Conference on Very Large Databases}, pages 610--621. Elsevier, 2002.

\bibitem{wishart2018drugbank}
David~S Wishart, Yannick~D Feunang, An~C Guo, Elvis~J Lo, Ana Marcu, Jason~R Grant, Tanvir Sajed, Daniel Johnson, Carin Li, Zinat Sayeeda, et~al.
\newblock Drugbank 5.0: a major update to the drugbank database for 2018.
\newblock {\em Nucleic acids research}, 46(D1):D1074--D1082, 2018.

\bibitem{johnson2016mimic}
Alistair~EW Johnson, Tom~J Pollard, Lu~Shen, Li-wei~H Lehman, Mengling Feng, Mohammad Ghassemi, Benjamin Moody, Peter Szolovits, Leo Anthony~Celi, and Roger~G Mark.
\newblock Mimic-iii, a freely accessible critical care database.
\newblock {\em Scientific data}, 3(1):1--9, 2016.

\bibitem{johnson2019mimic}
Alistair~EW Johnson, Tom~J Pollard, and Roger~G Mark.
\newblock Mimic-iii clinical database demo (version 1.4).
\newblock {\em PhysioNet}, 2019.

\bibitem{melnik2002similarity}
Sergey Melnik, Hector Garcia-Molina, and Erhard Rahm.
\newblock Similarity flooding: A versatile graph matching algorithm and its application to schema matching.
\newblock In {\em Proceedings 18th international conference on data engineering}, pages 117--128, 2002.

\bibitem{karras2020a}
Tero Karras, Samuli Laine, Miika Aittala, Janne Hellsten, Jaakko Lehtinen, and Timo Aila.
\newblock Analyzing and improving the image quality of stylegan.
\newblock In {\em Proceedings of the IEEE/CVF Conference on Computer Vision and Pattern Recognition (CVPR}, pages 8110–8119,, 2020.

\bibitem{subramanian2017a}
Sandeep Subramanian, Sai Rajeswar, Francis Dutil, Christopher Pal, and Aaron Courville.
\newblock Adversarial generation of natural language.
\newblock In {\em Proceedings of the 2nd Workshop on Representation Learning for NLP}, pages 241–251,, 2017.

\bibitem{choi2017a}
Edward Choi, Siddharth Biswal, Bradley Malin, Jon Duke, Walter~F. Stewart, and Jimeng Sun.
\newblock Generating multi-label discrete patient records using generative adversarial networks.
\newblock In {\em Machine learning for healthcare conference}, pages 286–305,, 2017.

\bibitem{park2018a}
Noseong Park, Mahmoud Mohammadi, Kshitij Gorde, Sushil Jajodia, Hongkyu Park, and Youngmin Kim.
\newblock Data synthesis based on generative adversarial networks.
\newblock {\em Proceedings of the VLDB Endowment}, 11(10):1071–1083, 2018.

\bibitem{mottini2018a}
Alejandro Mottini, Alix Lheritier, and Rodrigo Acuna-Agost.
\newblock Airline passenger name record generation using generative adversarial networks, 2018.
\newblock arXiv preprint arXiv:1807.06657,.

\bibitem{xu2019a}
Lei Xu, Maria Skoularidou, Alfredo Cuesta-Infante, and Kalyan Veeramachaneni.
\newblock Modeling tabular data using conditional gan.
\newblock In {\em Advances in Neural Information Processing Systems (NeurIPS}, volume~33. 2019.

\bibitem{koivu2020a}
Aki Koivu, Mikko Sairanen, Antti Airola, and Tapio Pahikkala.
\newblock Synthetic minority oversampling of vital statistics data with generative adversarial networks.
\newblock {\em Journal of the American Medical Informatics Association}, 27(11):1667–1674, 2020.

\bibitem{ma2020a}
Chao Ma, Sebastian Tschiatschek, Richard Turner, Jose~Miguel Hernandez-Lobato, and Cheng Zhang.
\newblock Vaem: a deep generative model for heterogeneous mixed type data.
\newblock In {\em Advances in Neural Information Processing Systems (NeurIPS}, volume~33. 2020.

\bibitem{vardhan2020a}
L.Vivek~Harsha Vardhan and Stanley Kok.
\newblock Generating privacy-preserving synthetic tabular data using oblivious variational autoencoders.
\newblock In {\em Proceedings of the Workshop on Economics of Privacy and Data Labor at the 37th International Conference on Machine Learning (ICML}, 2020.

\bibitem{darabi2021a}
Sajad Darabi and Yotam Elor.
\newblock Synthesising multi-modal minority samples for tabular data, 2021.
\newblock arXiv preprint arXiv:2105.08204,.

\bibitem{arik2021tabnet}
Sercan~{\"O} Arik and Tomas Pfister.
\newblock Tabnet: Attentive interpretable tabular learning.
\newblock In {\em Proceedings of the AAAI Conference on Artificial Intelligence}, volume~35, pages 6679--6687, 2021.

\bibitem{somepalli2021a}
Gowthami Somepalli, Micah Goldblum, C.Bayan~Bruss Avi~Schwarzschild, and Tom Goldstein.
\newblock Saint: Improved neural networks for tabular data via row attention and contrastive pre-training, 2021.
\newblock arXiv preprint arXiv:2106.01342,.

\bibitem{kossen2021a}
Jannik Kossen, Neil Band, Clare Lyle, Aidan Gomez, Tom Rainforth, and Yarin Gal.
\newblock Self-attention between datapoints: Going beyond individual input-output pairs in deep learning.
\newblock In {\em Advances in Neural Information Processing Systems}. 2021.

\bibitem{yin2020a}
Pengcheng Yin, Graham Neubig, Wen-tau Yih, and Sebastian Riedel.
\newblock Tabert: Pretraining for joint understanding of textual and tabular data.
\newblock In {\em Proceedings of the 58th Annual Meeting of the Association for Computational Linguistics}, page 8413–8426. Association for Computational Linguistics, 2020.

\bibitem{padhi2021a}
Inkit Padhi, Yair Schiff, Igor Melnyk, Mattia Rigotti, Youssef Mroueh, Pierre Dognin, Jerret Ross, Ravi Nair, and Erik Altman.
\newblock Tabular transformers for modeling multivariate time series.
\newblock In {\em ICASSP 2021-2021 IEEE International Conference on Acoustics, Speech and Signal Processing (ICASSP}, page 3565–3569. IEEE, 2021.

\bibitem{devlin2018a}
Jacob Devlin, Ming-Wei Chang, Kenton Lee, and Kristina Toutanova.
\newblock Bert: Pre-training of deep bidirectional transformers for language understanding, 2018.
\newblock arXiv preprint arXiv:1810.04805,.

\bibitem{borisov2022language}
Vadim Borisov, Kathrin Se{\ss}ler, Tobias Leemann, Martin Pawelczyk, and Gjergji Kasneci.
\newblock Language models are realistic tabular data generators.
\newblock {\em arXiv preprint arXiv:2210.06280}, 2022.

\bibitem{baowaly2019synthesizing}
Mrinal~Kanti Baowaly, Chia-Ching Lin, Chao-Lin Liu, and Kuan-Ta Chen.
\newblock Synthesizing electronic health records using improved generative adversarial networks.
\newblock {\em Journal of the American Medical Informatics Association}, 26(3):228--241, 2019.

\end{thebibliography}

\newpage

\section{Appendix}

\subsection{Prompt to generate base tables}

We provide an example to generate the instance of table ``Listed Security'' with model \texttt{text-davinci-003}. Here, we require the table contents to be generated conditional a list of given ``Legal Name'' column entries.

\begin{figure}[ht]
\begin{tcolorbox}[left=1mm,top=1mm,right=1mm,bottom=1mm]
\textbf{Input:} \\ 
Table Name: Listed Security  \\
Column Names: Ticker Symbol, Legal Name, Currency and Last Traded Value Monetary Amount \\
Dependent Columns: Column Currency depends on Currency of table Currency Name;
Last Traded Value Monetary Amount depends on Amount of table Monetary Amount;

\textbf{Prompt:} 
The task is to generate row examples for the table. 
For example, the table named Securities Transaction has columns: [`Type', `Count', `Settlement Date', `Facilitated By Financial Service Account', `Price Monetary Amount', `refers To Listed Security']. 
And column 'Facilitated By Financial Service Account' comes from the 'Account Number' of table 'Financial Service Account'. 
Column 'Price Monetary Amount' comes from 'Amount' of table Monetary Amount. 
Column 'refers To Listed Security' comes from 'Legal Name' of table 'Listed Security'. 
The row examples should look like 
'Example 1: Buy; 1020; 01/20/2022; 310689442; 200; Apple Inc.; Example 2: Sell; 638; 02/22/2023; 432689224; 138; Microsoft Inc.'. 
Now, the table named Listed Security have the following columns: 
['Ticker Symbol', 'Legal Name', 'Currency', 'Last Traded Value Monetary Amount']. 
The entries from column `Currency' comes from `Currency' of table `Currency Name'. 
The entries from 'Last Traded Value Monetary Amount' belong to `Amount' of table 'Monetary Amount'. 
\textit{Given the entries of column 'Legal Name' are ['Apple Inc.', 'Microsoft Inc.', 'Amazon Inc.', 'Alphabet Inc.', 'Facebook Inc.'].} Generate the rows for the table.

\textbf{Generated Contents:} \\
Example 1: AAPL; Apple Inc.; USD; 200   \\
Example 2: MSFT; Microsoft Inc.; USD; 138  \\
Example 3: GOOGL; Alphabet Inc.; USD; 150  \\
Example 4: AMZN; Amazon.com Inc.; USD; 250  \\
Example 5: FB; Facebook Inc.; USD; 100
\end{tcolorbox}

\caption{Example prompt generates values for the \emph{``listed security''} table in \Cref{fig:ontology-to-table}.}
\label{lst:prompt}
\end{figure}

\subsubsection{Algorithm to generate base tables}

Since the base tables depend on each other by the column correlations, simply generate the base tables independent could break the column correlations. The following algorithm enables the generated tables could satisfy the column correlation constrains.

\begin{algorithm}[ht]
\caption{Base Tables Generation with LLM}
\label{alg:algo1}
\begin{algorithmic}[1]
  \STATE \textbf{Input}: Column dependency relationships $R$ containing all pairs $((i,a), (j,b))$ satisfying column $a$ (from table $i$) depending on column $b$ (from table $j$).  
  \STATE \textbf{Build relation graph}: 
  \STATE For each table $i=1,2,...n$, create a node $v_i$.
  \STATE For all pairs $((i,a), (j,b)) \in R$, adding an edge $e_{ij}$ from node $v_i$ pointing to $v_j$. 
  \STATE \textbf{Generate tables with LLM}:
  \STATE Initialize generated tables set $Q = \emptyset$.
  \STATE For node $i$ with in degrees = 0, generate table $i$ with LLM. Insert $v_i$ to $Q$. 
  \WHILE{$|Q| < n$}
  \FOR{$j \notin Q$}
  \STATE Let $predecessors(v_j)$ be the set of all predecessors of $v_j$
  \IF{$predecessors(v_j) \subseteq Q$}
  \STATE Retrieve the columns that table $j$ depends on, namely $C_{i_1,a_1}, C_{i_2,a_2}, ..., C_{i_k,a_k}$
  \STATE Generate partitions of table $j$ with LLM conditional on the $C_{i_1,a_1}, C_{i_2,a_2}, ..., C_{i_k,a_k}$ respectively
  \STATE Concatenate the partitions into table $j$. Insert $v_j$ to $Q$
  \ENDIF
  \ENDFOR
  \ENDWHILE
\end{algorithmic}
\end{algorithm}

\subsection{Data Perturbation Functions}

Our pipeline also enables the following data perturbation functions on three different levels.

\definecolor{Mercury}{rgb}{0.901,0.901,0.901}
\definecolor{BlueRomance}{rgb}{0.843,0.968,0.831}
\definecolor{Onahau}{rgb}{0.8,0.964,1}
\definecolor{Pippin}{rgb}{1,0.901,0.898}
\begin{table}[ht]
\centering
\caption{\method{}'s tabular data perturbation functions.}
\label{tab:perturb_funcs}
\resizebox{\linewidth}{!}{%
\textscale{.8}{
\begin{tblr}{
  width = \linewidth,
  colspec = {Q[180]Q[240]Q[585]},
%   cells = {Pippin},
%   row{1} = {Mercury},
%   row{2} = {BlueRomance},
%   row{3} = {BlueRomance},
%   row{4} = {BlueRomance},
%   row{5} = {BlueRomance},
%   row{6} = {Onahau},
%   row{7} = {Onahau},
  cell{2}{1} = {r=4}{},
  cell{6}{1} = {r=2}{},
  cell{8}{1} = {r=5}{},
%   vlines = {Mercury},
  vline{2-3} = {1-12}{black},
  vline{3} = {1-12}{black},
  hline{1-2,13} = {-}{0.08em},
  hline{3-5,7,9-12} = {2-3}{Mercury},
  hline{6,8} = {-}{},
}
\textbf{} & \textbf{Perturbation Type }                                & \textbf{Perturbation action}                                                                                                             \\
\textbf{Column-values}      & Numeric noise                                     & Addition of jitter, duplicates                                                                                                              \\
                   & Text noise                                        & Typos, synonyms, abbreviations, word removal,                                                                                   \\
                   & Format noise                                      & Dates, addresses, names, etc.                                                                                                   \\
                   & Null-values                                       & Introduction of nulls                                                                                                           \\
\textbf{Schema}             & Cryptify column names                    & Cryptify business column names (e.g., abbreviate)                                                         \\
                   & Typos                                             & Introduce typos                                                                                                                 \\
\textbf{Table-shape}        & Removal of Columns                                & Remove columns from table                                                                                                                  \\
                   & Sampling                                          & Sample column contents                                                                                                                \\
                   & Vertical splitting                     & Split a table vertically with overlapping columns (10\%, 20\%, ...)                                                             \\
                   & Vertical splitting with unique column (PK/FK) & Split a table vertically with overlapping columns (10\%, 20\%, ...). One of the two split tables should only keep unique values. \\
                   & Horizontal Splitting                              & Horizontally split table with overlapping rows (0\%, 10\%, 20\%, ...)                                                                      
\end{tblr}}
}
\end{table}

\end{document}